\documentclass[conference]{IEEEtran}
\IEEEoverridecommandlockouts
\usepackage{cite}
\usepackage{amsmath,amssymb,amsfonts,amsthm}
\usepackage[theorems,skins]{tcolorbox}
\usepackage{algorithmic}
\usepackage{graphicx}
\usepackage{textcomp}
\usepackage{xcolor}
\usepackage{bm}
\usepackage{empheq}
\def\BibTeX{{\rm B\kern-.05em{\sc i\kern-.025em b}\kern-.08em
    T\kern-.1667em\lower.7ex\hbox{E}\kern-.125emX}}
\begin{document}

\title{Beam Pattern Modulation Embedded mmWave Hybrid Transceiver Design Towards ISAC
}

\author{\IEEEauthorblockN{ Boxun Liu$^*$, Shijian Gao$^\dagger$, Zonghui Yang$^*$, Xiang Cheng$^*$}
\IEEEauthorblockA{$^*$State Key Laboratory of Advanced Optical Communication
Systems and Networks \\
School of Electronics, Peking University, Beijing, China.\\
$^\dagger$Internet of Things Thrust, The Hong Kong University of Science and Technology (Guangzhou), Guangzhou, China.\\}
Email: \{boxunliu, yzh22\}@stu.pku.edu.cn,  shijiangao@hkust-gz.edu.cn, xiangcheng@pku.edu.cn
}

\maketitle

\begin{abstract}

Integrated Sensing and Communication (ISAC) emerges as a promising technology for B5G/6G, particularly in the millimeter-wave (mmWave) band. However, the widespread adoption of hybrid architecture in mmWave systems compromises multiplexing gain due to limited radio-frequency chains, resulting in mediocre performance when embedding sensing functionality. To avoid sacrificing the spectrum efficiency in hybrid structures while addressing performance bottlenecks in its extension to ISAC, we present an optimized beam pattern modulation-embedded ISAC (BPM-ISAC). BPM-ISAC applies index modulation over beamspace by selectively activating communication beams, aiming to minimize sensing beampattern mean squared error (MSE) under communication MSE constraints through dedicated hybrid transceiver design. Optimization involves the analog part through a min-MSE-based beam selection algorithm, followed by the digital part using an alternating optimization algorithm. Convergence and asymptotic pairwise error probability (APEP) analyses accompany numerical simulations, validating its overall enhanced ISAC performance over existing alternatives.

\end{abstract}


\section{Introduction}
Integrated Sensing and Communications (ISAC) \cite{cheng2023intelligent} is a pivotal technology for B5G/6G, striving for simultaneous improvement of communication and sensing through dedicated design \cite{fan2022radar}.
Recently, millimeter-wave (mmWave) ISAC has gained substantial attention due to the broader bandwidth it offers for both communication and sensing. The mmWave band's shared channel characteristics and signal processing techniques further facilitate integration.
However, the spectral efficiency (SE) of ISAC systems is often hindered by additional sensing functions. This issue is aggravated in hybrid mmWave systems, where limited radio-frequency (RF) chains severely compromise the multiplexing gain.

Index modulation (IM) \cite{younis2010generalised,gao2019spatial} emerges as a promising technology to address the aforementioned issue. However, most IM-embedded ISAC beamforming designs \cite{huang2020majorcom,ma2021spatial,xu2022hybrid}, relying on antenna activation-based spatial modulation, are exclusively designed for fully digital (FD) architecture, limiting direct application to hybrid systems.
Although generalized beamspace modulation (GBM) \cite{gao2019spatial} can elevate SE by implementing IM over beamspace and remains compatible with mmWave hybrid structures, it solely focuses on communication without integrating sensing functions.

Moving from GBM, recent works \cite{elbir2023millimeter,elbir2023spatial} introduce spatial path index modulation into mmWave ISAC systems (SPIM-ISAC) to attain higher SE. 
This approach selectively activates partial spatial paths for communication and employs a single fixed beam for sensing. 
SPIM, a subset of GBM, constructs beamspace through fixed strongest channel paths without performance optimization. 
However, as it extends to multi-angle scanning and non-line-of-sight (NLoS) scenarios, the randomness of sensing angles introduces potential disturbance to communication users due to sensing beams. 
Additionally, SPIM-ISAC achieves a performance trade-off through power allocation between optimal communications-only and sensing-only beamformers, lacking a comprehensive consideration of overall performance.

To preserve the SE benefits of the original beamspace modulation while addressing current performance limitations in its extension to ISAC, we present an optimized beam pattern modulation-embedded ISAC (BPM-ISAC). 
BPM-ISAC generates multiple beams for single-user communication and scanning beams for sensing. 
To enhance sensing performance while ensuring communication reliability, we formulate a joint hybrid transceiver design problem aimed at minimizing the mean squared error (MSE) of the desired sensing beampattern under the constraint of communication MSE.
Optimal communication beam pairs are selected from the Discrete Fourier Transform (DFT) codebook, and the digital part is optimized using the proposed alternating optimization algorithm for improved power allocation, with proven convergence. 
Theoretical analyses of the distribution of effective paths and the asymptotic pairwise error probability (APEP) for BPM-ISAC are provided. 
Simulations and analyses collectively demonstrate that the proposed BPM-ISAC outperforms existing alternatives, offering overall improved ISAC performance.

\begin{figure*}[t]
\center{\includegraphics[width=16cm]  {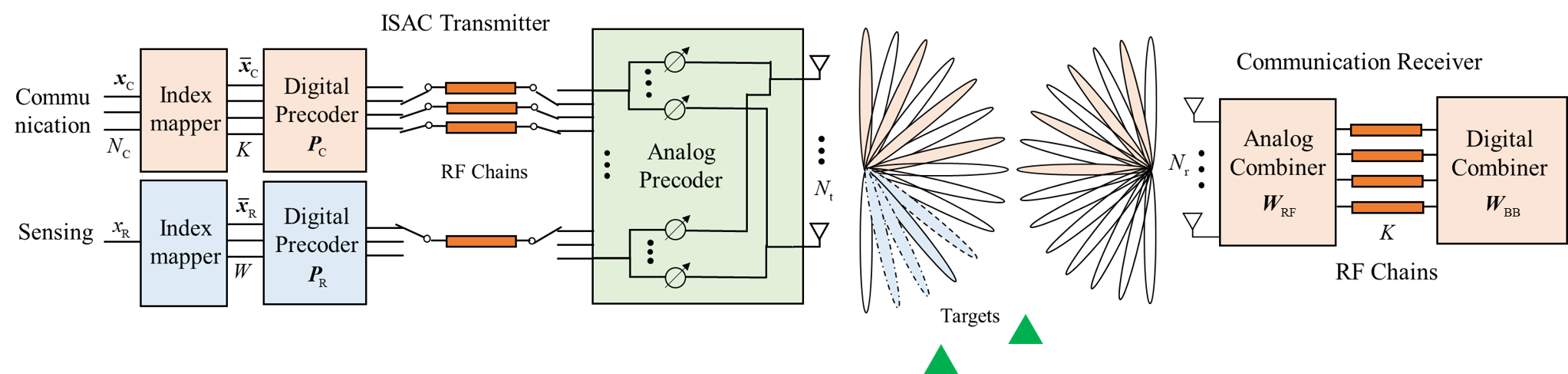}} 
\caption{Illustration of the transceiver architecture for the proposed BPM-ISAC mmWave system. ($N_{\rm C}=3$,$K=4$,$W=3$)}
 \label{System}
 \vspace{-5mm}
\end{figure*}

\textit{Notation}: $(\cdot)^{\rm T}$, $(\cdot)^{\rm H}$,  $\Vert\cdot\Vert_2$, $\Vert\cdot\Vert_F$ denote the transpose, the conjugate transpose, 2 norm, and  Frobenius norm, respectively.
$\mathcal{CN}(m,\sigma^2)$ represents the complex Gaussian distribution whose mean is $m$ and covariance is $\sigma^2$. 
$\bm{I}_K$ denotes the $K\times K$ identity matrix and $\bm{1}_K$ denotes the $K\times 1$ all-one column vectors.
${\rm{diag}}(\bm{a})$ denotes a diagonal matrix formed from vector $\bm{a}$ and $\mathbb{E}[\cdot]$ denotes the expectation operation.
\section{System Model and Problem Formulation}
In this paper, we consider a fully-connected hybrid mmWave ISAC multi-beam system for point-to-point communication and multi-target detection.
As shown in Fig. \ref{System}, the ISAC transmitter and communication receiver employ $N_{\rm t}$ and $N_{\rm r}$-antenna half-wavelength spaced uniform linear array, respectively.
\subsection{Symbol Modulation}
At the ISAC transmitter, $K$ communication beams and $W$ sensing scanning beams are generated using corresponding digital and analog precoders, i.e.,
\begin{align}
    \bm{s}=\bm{F}_{\rm C}\bm{P}_{\rm C}\bar{\bm{x}}_{\rm C}+\bm{F}_{\rm R}\bm{P}_{\rm R}\bar{\bm{x}}_{\rm R}.
\end{align}
$\bm{\bar{x}}_{\rm C} \in \mathbb{C}^{K \times 1}$ and $\bm{\bar{x}}_{\rm R}\in\mathbb{C}^{W \times 1}$ are the modulated communication and sensing symbols, respectively. 
$\bm{F}_{\rm C}\in\mathbb{C}^{N_{\rm t}\times K}$  and $\bm{F}_{\rm R}\in\mathbb{C}^{N_{\rm t}\times W}$ are analog precoders for communication and sensing, respectively, while $\bm{P}_{\rm C}={\rm{diag}}\left(\bm{p}\right)\in\mathbb{R}^{K \times K}$ and $\bm{P}_{\rm R}={\rm{diag}}\left(\bm{b}\right)\in\mathbb{R}^{W \times W}$ are their associated digital precoders for power allocation across beams.

For communication, we apply IM on the beam domain, i.e., $N_{\rm C}$ out of $K$ beams are activated simultaneously. 
Hence the $N_{\rm C}$-dimensional non-zero data stream $\bm{x}_{\rm C} \in \mathbb{C}^{N_{\rm C}}$ is modulated to $K$-dimensional zero-containing $\bm{\bar{x}}_{\rm C}$ with totally ${\rm C}_K^{N_{\rm C}}$ possible index patterns.
Similar to \cite{gao2019spatial}, $2^{\lfloor {\rm log_{2}}{{\rm C}_K^{N_{\rm C}}} \rfloor}$ of these patterns are utilized to transmit additional  $\lfloor {\rm log_{2}}{{\rm C}_K^{N_{\rm C}}} \rfloor$ index bits. Suppose $\bm{x}_{\rm C}$ adopts M-ary phase shift keying/quadrature amplitude modulation (PSK/QAM) and the SE is given by
\begin{align}
    \eta=N_{\rm C}\log_2{M}+\lfloor{\rm log_{2}}{{\rm C}_K^{N_{\rm C}}} \rfloor {\rm bps/Hz}.
\end{align}

For sensing, to minimize the use of RF chains, each of the $W$ beams is sequentially activated to scan $W$ directions of interest. 
Therefore, sensing signal $x_{\rm R}$ is mapped to a $W$-dimensional one-hot vector $\bm{\bar{x}}_{\rm R}\in\mathbb{C}^{W \times 1}$ before transmission.
To achieve more adjustable detection, the case of non-equal probability scanning is considered.
Denote the activation probability matrix as $\bm{D}={\rm{diag}}([d_1,...,d_W])$, where $d_i$ represents the activation probability of $i\mbox{-}$th sensing beam.
\subsection{Channel Model and 
Communication MSE}
We adopt classical Saleh-Valenzuela narrow-band mmWave channel \cite{saleh1987statistical} with $P$ dominant paths
\begin{align}\label{channel} 
	\bm{H}=\sqrt{\frac{N_{\rm t}N_{\rm r}}{P}}\sum_{i = 1}^{P}\alpha_i\bm{a}_{\rm r}\left(\theta_i\right)\bm{a}_{\rm t}^{\rm H}\left(\phi_i\right), 
\end{align}
where $\alpha_i\sim\mathcal{CN}(0,1)$ is the gain of $i\mbox{-}$th path, and its angles of arrival and departure (AoAs/AoDs) $\theta_i$ and $\phi_i$ are uniformly distributed in $[-\pi/2,\pi/2)$.
$\bm{a}_{\rm r}\left(\theta_i\right)$ and $\bm{a}_{\rm t}\left(\phi_i\right)$ are the array steering vectors and 
\begin{align} \label{steer}
    \bm{a}(\theta)=\frac{1}{\sqrt{N_{\rm t}}}[1,e^{j\pi \sin(\theta)},...e^{j\pi(N_{\rm t}-1) \sin(\theta)}]^{\rm H}.
\end{align}
The received signal after analog combiner $\bm{W}_{\rm RF}\in\mathbb{C}^{N_{\rm r}\times K}$ becomes
\begin{align}
    \bm{y}_C =\bm{H}_{\rm C}\bm{P}_{\rm C}\bm{\bar{x}}_{\rm C}+\bm{H}_{\rm R}\bm{P}_{\rm R}\bm{\bar{x}}_{\rm R}+\bm{\xi}_{\rm C}, 
\end{align}
where $\bm{H}_{\rm C}=\bm{W}^{\rm H}_{\rm RF}\bm{H}\bm{F}_{\rm C}$ and $\bm{H}_{\rm R}=\bm{W}^{\rm H}_{\rm RF}\bm{H}\bm{F}_{\rm R}$ are the equivalent digital channel (EDC) for communication and sensing, and $\bm{\xi}_{\rm C} \sim \mathcal{CN}(0,\sigma^2 \bm{I}_K)$ is additive white Gaussian noise.
The LMMSE digital combiner \cite{fan2021wideband} is adopted as
\begin{align} \label{m}
\bm{W}_{\rm BB}=&\bm{R}_{\bm{\bar{x}}_{\rm C}}\bm{P}_{\rm C}\bm{H}_{\rm C}^{\rm H}\left(\bm{H}_{\rm C}\bm{P}_{\rm C}\bm{R}_{\bm{\bar{x}}_{\rm C}}\bm{P}_{\rm C}\bm{H}_{\rm C}^{\rm H}+\right.\nonumber\\
&\left.\bm{H}_{\rm R}\bm{P}_{\rm R}\bm{R}_{\bm{\bar{x}}_{\rm R}}\bm{P}_{\rm R}\bm{H}_{\rm R}^{\rm H}+\sigma^2\bm{I}_K\right)^{-1},
\end{align}
where 
$\bm{R}_{\bm{\bar{x}}_{\rm C}}=\mathbb{E}\left[\bm{\bar{x}}_{\rm C}\bm{\bar{x}}_{\rm C}^{\rm H}\right]=\frac{N_{\rm C}}{K}\bm{I}_K$ and $R_{\bm{\bar{x}}_{\rm R}}=\mathbb{E}\left[\bm{\bar{x}}_{\rm R}\bm{\bar{x}}_{\rm R}^{\rm H}\right]=\bm{D}$.
Then the symbol $\bm{\bar{x}}_{\rm C}$ is estimated as   
 \begin{align}\label{esti sym} \tilde{\bm{x}}_C=\bm{W}_{\rm BB}\left(\bm{H}_{\rm C}\bm{P}_{\rm C}\bm{\bar{x}}_{\rm C}+\bm{H}_{\rm R}\bm{P}_{\rm R}\bm{\bar{x}}_{\rm R}+\bm{\xi}_{\rm C}\right).
 \end{align}
Then the information bits contained in $\bm{x}_{\rm C}$ and index bits can be estimated by the maximum likelihood (ML) detector, which is omitted here.
According to Eq. (\ref{esti sym}), the symbol MSE of $\bm{\bar{x}}_{\rm C}$ denoted as  $\mathbb{E}\left(\Vert\tilde{\bm{x}}_C-\bm{\bar{x}}_{\rm C}\Vert_F^2\right)$, is derived as
\begin{align}\label{MSE}\chi
    =&\frac{N_{\rm C}}{K}\Vert\bm{W}_{\rm BB}\bm{H}_{\rm C}\bm{P}_{\rm C}-\bm{I}\Vert_F^2\nonumber\\
    &+\Vert\bm{D}^\frac{1}{2}\bm{W}_{\rm BB}\bm{H}_{\rm R}\bm{P}_{\rm R}\Vert_F^2 +\sigma^2\Vert\bm{W}_{\rm BB}\Vert_F^2.
\end{align}
Due to the challenges of establishing appropriate suitable MSE constraints in practical issues due to different channel conditions, a relative MSE threshold is introduced as

 \begin{align}\label{threshold}
     \Gamma(\mu)=&\frac{N_{\rm C}}{K}\Vert\bm{W}_{\rm BB,0}\bm{H}_{\rm C}-\bm{I}_K\Vert_F^2+\nonumber\\
     &\mu\Vert\bm{D}^\frac{1}{2}\bm{W}_{\rm BB,0}\bm{H}_{\rm R}{\rm{diag}}(\bm{t})\Vert_F^2+\sigma^2\Vert\bm{W}_{\rm BB,0}\Vert_F^2,
 \end{align}
where 
${\bm{W}_{\rm BB,0}\hspace{-0.04in}=\hspace{-0.04in}\frac{N_{\rm C}}{K}\bm{H}_{\rm C}^{\rm H} (\frac{N_{\rm C}}{K}\bm{H}_{\rm C}\bm{H}_{\rm C}^{\rm H}\hspace{-0.04in}+\hspace{-0.04in}\bm{H}_{\rm R}\left({\rm{diag}}\left(\bm{t}\right)\right)^2\bm{D}\bm{H}_{\rm R}^{\rm H}}+\sigma^2\bm{I}_K)^{-1}$.
$0\le\mu\le1$ is a weighting coefficient, signifying the relative tolerance for sensing interference cancellation errors.
\subsection{Sensing Beampattern MSE}
Sensing beampattern MSE \cite{cheng2022qos} is a crucial metric ensuring sufficient radiation energy in the directions of interest.
Denote the $i\mbox{-}$th element of $\bm{b}$ as $b_i$, standing for the allocated power on the $i\mbox{-}$th sensing beam.
Neglecting the energy of communication beams, the sensing beampattern is defined as
\begin{align}
\bm{v}=[\lvert b_1 \bm{a}^{\rm H}(\theta_1){\bm{F}_{\rm R}[:,1]}\rvert,..,\lvert b_W \bm{a}^{\rm H}(\theta_W)\bm{F}_{\rm R}[:, W]\rvert]^{\rm T},
\end{align}
where $\theta_t$ denotes the $t\mbox{-}$th direction of interest.
Assuming a desired beampattern $\bm{t}\in \mathbb{R}^W$, it should satisfy the power constraint $\Vert \bm{D}^\frac{1}{2}\bm{t}\Vert_2^2=\sum_{i=1}^{W}d_it_i^2=T_{\rm R}$, where $t_i$ is the $i\mbox{-}$th element of $\bm{t}$ and $T_{\rm R}$ is the average sensing power.
The sensing beampattern MSE is derived as
\begin{align}
 \Vert \bm{D}^\frac{1}{2}(\bm{v}-\bm{t})\Vert_2^2=\sum_{i=1}^{W}d_i(v_i-t_i)^2.
\end{align}
\vspace{-5mm}
\subsection{Problem Formulation}
To achieve the desired sensing beampattern with guaranteed communication reliability, our goal is to minimize the MSE of the sensing beampattern while adhering to constraints on communication MSE, transmit power, and analog precoder.
The optimization problem is formulated as follows:
\vspace{-2mm}
\newtcolorbox{mymathbox}[1][]{colback=white, sharp corners, boxrule={0.2 mm},top={0mm},boxsep={0mm},height={48mm}, toptitle={0.1mm},bottom={0 mm}}
\begin{mymathbox}[ams equation]
\begin{subequations}
\vspace{-3mm}
\begin{align}
\mathcal{P}.1:~&\textbf{Hybrid transceivers designed for ISAC}\nonumber\\
\min_{\substack{\bm{P}_{\rm C},\bm{P}_{\rm R}\\ \bm{F}_{\rm C},\bm{F}_{\rm R},\bm{W}_{\rm RF}}} &\Vert\bm{D}^\frac{1}{2}(\bm{v}-\bm{t})\Vert_2^2\nonumber\\
 s.t. \quad\enspace\, &\chi \le \Gamma(\mu), \label{MSE Cons}\\
 &\Vert{\rm{diag}}(\bm{p})\Vert_F^2\le K,\label{comm power}\\
 &\Vert\bm{D}^{\frac{1}{2}}\bm{b}\Vert_F^2\le T_{\rm R}, \label{radar power0}\\ &\bm{F}_{\rm C}\in\mathcal{F}, \label{ct analog set}
 \\ &\bm{F}_{\rm R}\in\mathcal{F}, \label{r analog set}\\
 &\bm{W}_{\rm RF}\in\mathcal{W}.\label{cr analog set}
\end{align}
\end{subequations}
\end{mymathbox}
\noindent
The expressions of $\chi$ and $\Gamma(\mu)$ are provided in Eqs. (\ref{MSE}) and (\ref{threshold}), respectively. 
(\ref{MSE Cons}) represents MSE constraint on communication symbol estimation, (\ref{comm power}) corresponds to the communication power constraint, and (\ref{radar power0}) denotes the average sensing power constraint. 
In Eqs. (\ref{ct analog set})-(\ref{cr analog set}), $\mathcal{F}$ and $\mathcal{W}$ represent the feasible sets of analog precoders and combiners.
\section{Hybrid Transceivers Optimization}
Problem $\mathcal{P}.1$ is a complex non-convex optimization problem that is barely tractable. 
Therefore, we solve it by optimizing the analog and digital parts sequentially.
\subsection{Analog-Part Optimization}
Firstly, the analog part is optimized to build EDC with the unoptimized digital part, i.e., $\bm{P}_{\rm R}={\rm{diag}}(\bm{t})$,  $\bm{P}_{\rm C}=\bm{I}_K$, $\bm{W}_{\rm BB}=\bm{W}_{\rm BB,0}$.
In this paper, we adopt the commonly used DFT codebook to construct the analog part.
Specifically, each column of $\bm{F}_{\rm C}$, $\bm{F}_{\rm R}$, and $\bm{W}_{\rm RF}$ is selected from $\mathcal{F}_{N}=\{\bm{f}(1),...,\bm{f}(N)\}$, where $\bm{f}(i)=\frac{1}{N_{\rm t}}[1,e^{j2\pi\frac{(i-1)}{N_{\rm t}}},...,e^{j2\pi\frac{(N_{\rm t}-1)(i-1)}{N_{\rm t}}}]^T$ and $N$ takes on $N_{\rm t}$ or $N_{\rm r}$.

For sensing, we assume that the directions of interest fall within the directions of DFT codewords, which is reasonable for massive antennas. 
Thus the sensing beams can accurately point to  $W$ target directions, and $\bm{F}_{\rm R}[:,l]=\bm{a}(\theta_l)$ is satisfied for all $l$.
Considering the sensing beam set as $\Omega=\{\bm{F}_{\rm R}[:,1],...,\bm{F}_{\rm R}[:,W]\}$,  communication transmitting beams should be selected from $\mathcal{F}_{N_{\rm t}}\backslash
\Omega$ to avoid sensing interference on the communication receiver.
For communication, the analog precoder and combiner are optimized to minimize communication MSE, following as
\begin{align}
    \bar \chi=&\frac{N_{\rm C}}{K}\Vert\bm{W}_{\rm BB,0}\bm{H}_{\rm C}-\bm{I}_K\Vert_F^2+\nonumber\\
    &\Vert\bm{D}^\frac{1}{2}\bm{W}_{\rm BB,0}\bm{H}_{\rm R}{\rm{diag}}(\bm{t})\Vert_F^2+\sigma^2\Vert\bm{W}_{\rm BB,0}\Vert_F^2.
\end{align}
Accordingly, the optimal communication beam pairs are selected from the DFT codebook based on the min-MSE criterion, i.e.,
\begin{align}
    \left\{\bm{F}_{\rm C}, \bm{W}_{\rm RF}\right\}=\mathop{\arg\min}\limits_{
    \substack{\forall i, \bm{F}_{\rm C}[:,i]\in\mathcal{F}_{N_{\rm t}}\backslash
\Omega\\\bm{W}_{\rm RF}[:,i]\in\mathcal{F}_{N_{\rm r}}}}\bar \chi.
\end{align}
Considering the exponential time complexity of the exhaustive search, we proposed a two-stage alternative to lower complexity.
In the first stage, we obtain a set of $L$ strongest candidate beam pairs $\{(\bm{f}(n_i),\bm{f}(m_i)),i=1,..., L\}$ based on a max-power criterion, where the power of beam pairs is defined as $\bm{f}^{\rm H}(m_i)\bm{H}\bm{f}(n_i)$. 
In the second stage, the final beam pairs are selected from the $L$ candidate beam pairs using the min-MSE criterion.
In this manner, the complexity will be reduced from $\mathcal{O}(C_{N_{\rm r}}^K C_{N_{\rm t}-W}^K)$ to $\mathcal{O}((N_r\times (N_t-W))^2+C_L^K)$.
\subsection{Digital-Part Optimization}
The digital part is optimized with fixed EDC as follows:
\begin{subequations}\label{digital}
\begin{empheq}[box=\fbox]{align}
\mathcal{P}.2:~&\textbf{Beam power allocation}\nonumber\\
&\min_{\substack{\bm{b},\bm{p}\\ }} \enspace \Vert\bm{D}^\frac{1}{2}(\bm{b}-\bm{t})\Vert_2^2 \label{digital opject}\nonumber\\
 & \, \,s.t.\quad \left(\ref{MSE Cons}\right)-\left(\ref{radar power0}\right).\nonumber
\end{empheq}
\end{subequations}
Since $\bm{b}$, $\bm{p}$ and $\bm{W}_{\rm BB}$ are coupled in Eq. (\ref{m}) in a non-convex manner, $\mathcal{P}.2$ is a non-convex problem.
In the following steps, an alternating optimization algorithm is proposed.
For initialization, we set $\bm{b}$, $\bm{p}$, and $\bm{W}_{\rm BB}$ as $\bm{t}$, $\bm{1}_K$, and $\bm{W}_{\rm BB,0}$, respectively.
In each iteration, $\bm{b}$, $\bm{p}$ and $\bm{W}_{\rm BB}$ are optimized sequentially.

\noindent
1) Update $\bm{b}$ with fixed $\bm{p}$ and $\bm{W}_{\rm BB}$ by solving
\begin{subequations}
    \begin{align}
\min_{\bm{b}}&\Vert\bm{D}^\frac{1}{2}(\bm{v}-\bm{t})\Vert_2^2\nonumber\\
      s.t.& \quad (\ref{MSE Cons}) (\ref{radar power0}).\nonumber
    \end{align}
\end{subequations}

\noindent
2) Update $\bm{p}$ with fixed $\bm{b}$ and $\bm{W}_{\rm BB}$ by solving
\begin{subequations}\label{find p}
\begin{align}
    \min_{\bm{p}} & \quad \frac{N_{\rm C}}{K}\Vert\bm{W}_{\rm BB}\bm{H}_{\rm C}{\rm{diag}}(\bm{p})-\bm{I}\Vert_F^2+\nonumber\\
    &\Vert\bm{D}^\frac{1}{2}\bm{W}_{\rm BB}\bm{H}_{\rm R}{\rm{diag}}(\bm{b})\Vert_F^2+\sigma^2\Vert\bm{W}_{\rm BB}\Vert_F^2 \nonumber\\
    s.t.\,& \quad  (\ref{comm power}). \nonumber
\end{align}
\end{subequations}
3) Update $\bm{W}_{\rm BB}$ with fixed $\bm{b}$ and $\bm{p}$ according to  Eq. (\ref{m}).

It is worth noting that the problems in steps 1) and 2) are convex quadratically constrained quadratic programming (QCQP) problem and can be solved using the existing convex optimization toolbox \cite{lofberg2004yalmip}. 
The time complexity is
$\mathcal{O}((W^{3.5}+K^{3.5}){\rm log}(1/\epsilon))$ \cite{luo2010semidefinite} solving it by the interior-point method with an accuracy level $\epsilon$.
\vspace{-3mm}
\section{Performance Analysis}
\vspace{-1mm}
\subsection{Convergence Analysis}
Use superscripts to represent variable names for the $i\mbox{-}$th iteration. 
For the fist iteration, $\bm{b}^{(1)}$ must have a solution since $\bm{b}^{(1)}=\mu \bm{t}$ is a feasible solution when $0\le\mu\le1$. 
Similarly, $\bm{p}^{(1)}$ also has a solution since $\bm{p}^{(1)}=\bm{p}^{(0)}$ is feasible. 
Consider the $(i+1)\mbox{-}$th iteration and denote the objective function at the step $j$  as $\varepsilon_j$.
For step 1), $\varepsilon_1(\bm{b}^{(i+1)})\le\varepsilon_1(\bm{b}^{(i)})$. 
For step 2) and 3), $\varepsilon_3(\bm{b}^{(i+1)})=\varepsilon_2(\bm{b}^{(i+1)})=\varepsilon_1(\bm{b}^{(i+1)})$. Notice that constraint (\ref{MSE Cons}) is satisfied since
$\bm{p}^{(i+1)}$ is optimized to lower MSE and  $\bm{W}_{\rm BB}^{(i+1)}$ is the LMMSE equalizer to minimize MSE. Therefore, after the $(i+1)\mbox{-}$th iteration, the objective function is non-increasing, and all constraints are satisfied. Considering the objective value has a lower bound, the alternating optimization is guaranteed to converge.
\vspace{-3mm}
\subsection{APEP Analysis}
\begin{figure}[tbp]
    \centering    \includegraphics[scale=0.25]{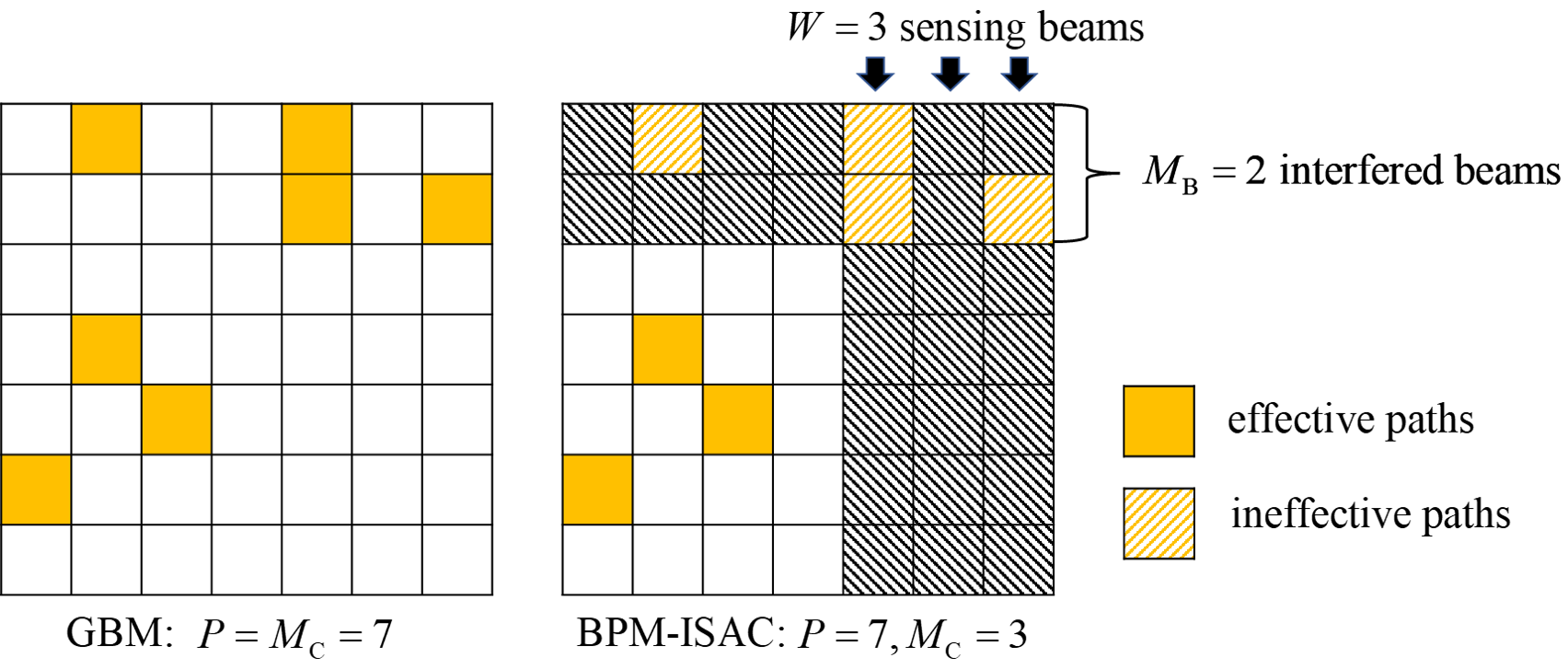}
    \vspace{-3mm}
    \caption{Illustration of beamspace channels for GBM and BPM-ISAC.}
    \label{effective paths}
    \vspace{-7mm}
\end{figure}
We conduct APEP analysis without optimizing the digital part since it applies the post power allocation without influencing the beam selection.
We assume infinite sensing interference power and neglect off-grid beam leakage.
Unlike GBM, the number of effective paths decreases due to the interference of sensing beams.
As shown in Fig. \ref{effective paths}, there are $P=7$ paths in the original $N_{\rm r} \times N_{\rm t}=7\times7$ beamspace. 
The $W=3$ sensing beams cover $M_{\rm R}=3$ paths, which further interferes with $M_{\rm B}=2$ received beams.
Thus only $M_{\rm C}=3$ effective paths are left in the undisturbed $(N_{\rm r}-M_{\rm B})\times(N_{\rm t}-W)$ sub-beamspace.
It can be derived that the probability that sensing beams cover $M_{\rm R}=r$ paths is 
\begin{align}\label{P_MR}
 P_{M_{\rm R}}(r)=\frac{C_{N_{\rm r}(N_{\rm t}-W)}^{P-r}C_{N_{\rm r}W}^{r}}{C_{N_{\rm t}N_{\rm r}}^P}, 
\end{align}
the probability that these paths cover $M_{\rm B}=b$ received beams is 
\begin{align}\label{P_MB}
P_{M_{\rm B}}(r,b)=
\left\{\begin{array}{l}P_{M_{\rm B}}(r-1,b-1)\frac{(N_{\rm r}-b+1)W}{N_{\rm r}W-r+1}\\
\qquad +P_{M_{\rm B}}(r-1,b)\frac{bW-r+1}{N_{\rm r}W-r+1}, o.w.
\\0, \quad \qquad (r,b)=(1,1) \quad or \quad b=0.
\end{array}\right.
\end{align}
Given $M_{\rm R}=r$ and $M_{\rm B}=b$, the probability that $M_{\rm C}=c$ paths are available for communication is
\begin{align}\label{P_MC}
P_{M_{\rm C}}(c, r, b)=\frac{C_{b(N_{\rm t}-W)}^{P-r-c}C_{(N_{\rm r}-r)(N_{\rm t}-W)}^c}{C_{N_{\rm r}(N_{\rm t}-W)}^{P-r}}.
\end{align}
Therefore, the probability distribution of the number of effective paths is derived as
\begin{align}\label{P com}
P(M_{\rm C}=c)=
\left\{\begin{array}{l} P_{M_{\rm R}}(0), c=P,
\\ \sum\limits_{r=1}^{P-c} \sum\limits_{b=1}^{r}P_{M_{\rm R}}(r)P_{M_{\rm B}}(r,b)P_{M_{\rm C}}(c,r,b), 
\\\qquad \qquad \qquad \qquad c=0,...,P-1.
\end{array}\right.
\end{align}
According to \cite{gao2019spatial}, the APEP is expressed as
 \begin{align}\label{APEP}
 P_{APEP}=&\frac{1}{\eta 2^\eta}\sum_{\bm{\bar{x}}_{\rm C}}\sum_{\bm{\hat{x}}_C}P(\bm{\bar{x}}_{\rm C} \rightarrow \bm{\hat{x}}_C)e(\bm{\bar{x}}_{\rm C}, \bm{\hat{x}}_C),
 \end{align}
where $P(\bm{\bar{x}}_{\rm C} \rightarrow \bm{\hat{x}}_C)$ represents the pairwise error probability and
$e(\bm{\bar{x}}_{\rm C}, \bm{\hat{x}}_C)$  
represents the number of error bits between $\bm{\bar{x}}_{\rm C}$ and $\bm{\hat{x}}_C$, respectively. 
For $M_{\rm C}<K$, we assume that a BER of 0.5 in this case.
For $M_{\rm C}\geq K$, it is equivalent to GBM case with $c$ paths.
Hence $P(\bm{\bar{x}}_{\rm C} \rightarrow \bm{\hat{x}}_C)$ is derived as
\begin{align}\label{Q_f}
P(\bm{\bar{x}}_{\rm C} \hspace{-0.04in}\rightarrow \hspace{-0.04in}\bm{\hat{x}}_C)\hspace{-0.04in}=\hspace{-0.04in}\sum\limits_{c=K}^P \hspace{-0.04in} P(M_{\rm C}=c)
P_c(\bm{\bar{x}}_{\rm C},\bm{\hat{x}}_C)\hspace{-0.04in}+\hspace{-0.04in}\frac{1}{2^\eta}P(M_{\rm C}<K),
\end{align}
where $P_c(\bm{\bar{x}}_{\rm C},\bm{\hat{x}}_C)$ denotes the pairwise error probability with $c$ effective paths.
Referring to \cite{fan2021wideband}, it can be derived that
\begin{align}
P_c(\bm{\bar{x}}_{\rm C},\bm{\hat{x}}_C)&=\frac{\mathbb{B}\left(\sum_{i=1}^{K}\left(\frac{N_{\rm t}N_{\rm r}}{4P\sigma^2}{\bigtriangleup x_i^2}+1\right),c-K+1\right)}{12\prod \limits_{j=2}^{K}\sum_{i=j}^{K}\left(\frac{N_{\rm t}N_{\rm r}}{4P\sigma^2}{\bigtriangleup x_i^2}+1\right)}+
\nonumber\\
&\frac{\mathbb{B}\left(\sum_{i=1}^{K}\left(\frac{N_{\rm t}N_{\rm r}}{3P\sigma^2}{\bigtriangleup x_i^2}+1\right),c-K+1\right)}{4\prod \limits_{j=2}^{K}\sum_{i=j}^{K}\left(\frac{N_{\rm t}N_{\rm r}}{3P\sigma^2}{\bigtriangleup x_i^2}+1\right)}, 
\end{align}
where $\bigtriangleup x_i=\bm{\bar{x}}_{\rm C}[i]-\bm{\hat{x}}_C[i]$.
\vspace{-1mm}
\section{Simulations}
\vspace{-2mm}
\begin{figure}[t]
    \centering
    \includegraphics[scale=0.5]{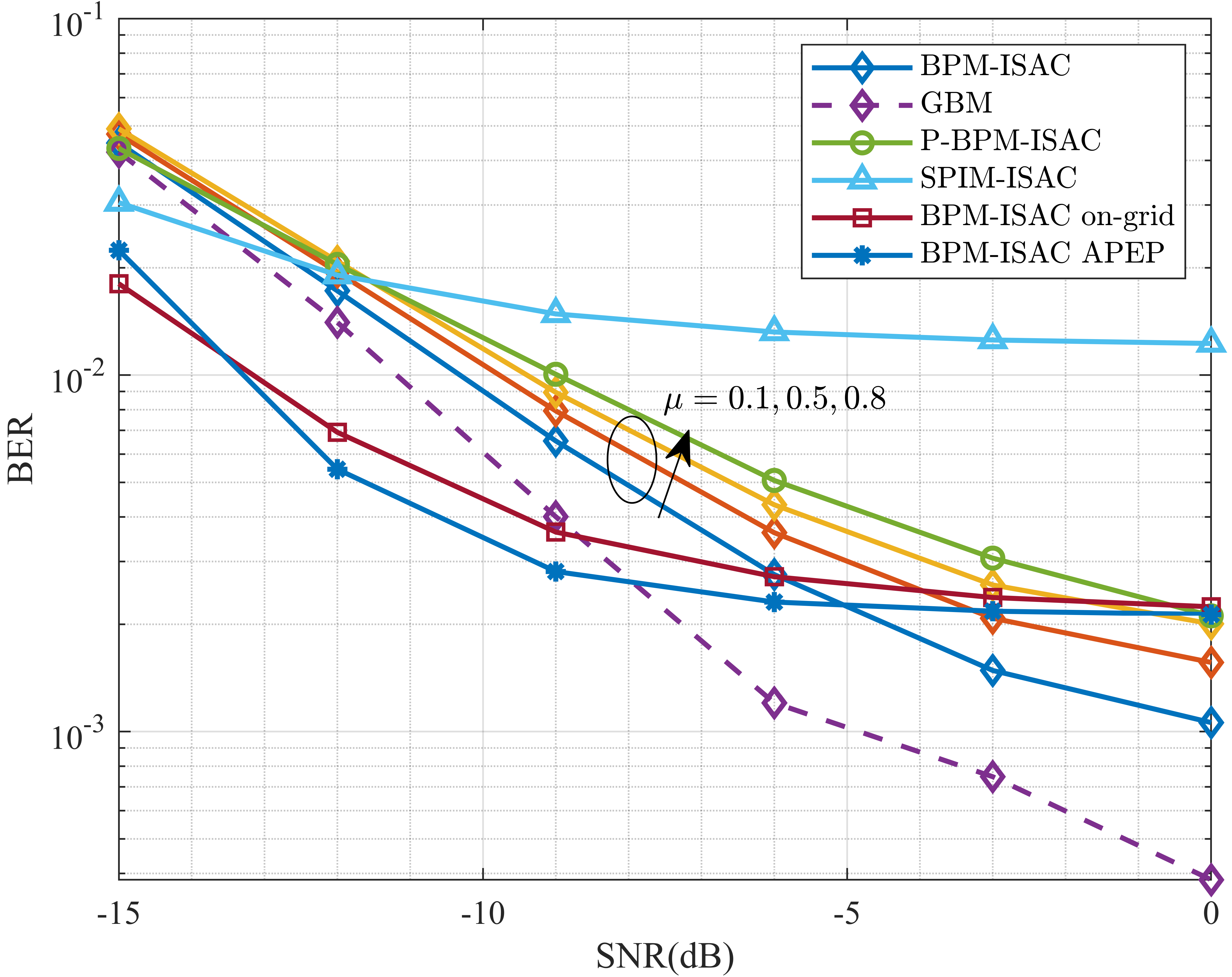}
    \vspace{-2mm}
    \caption{BER comparison among different methods.}
    \label{BER}
    \vspace{-7mm}
\end{figure}
\begin{figure}[t]
    \centering
    \includegraphics[scale=0.3]{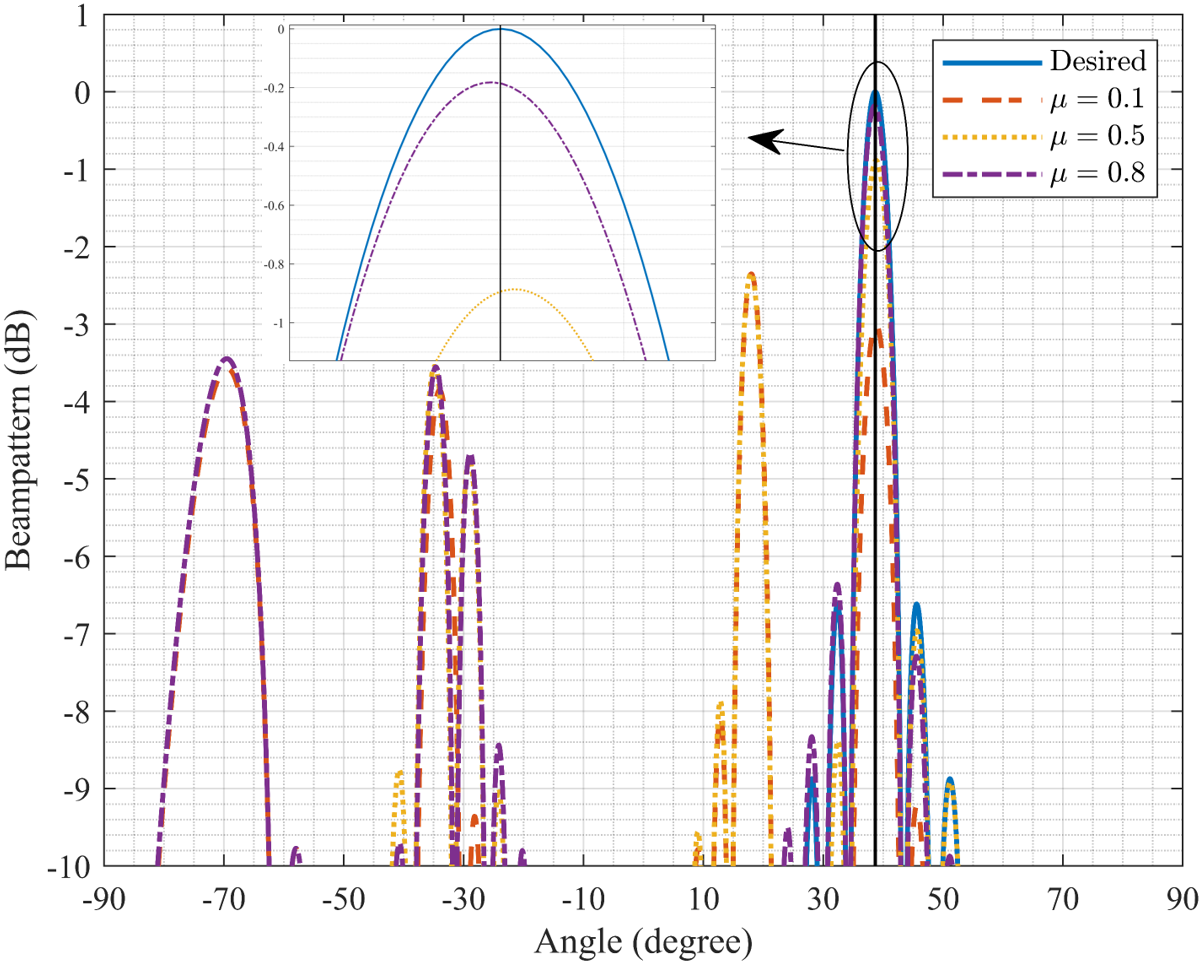}
    \vspace{-2mm}
    \caption{Illustration of the normalized beampattern under different weighting coefficient $\mu$.}
    \label{beampattern}
    \vspace{-5mm}
\end{figure}
\begin{figure}[t]
    \centering    \includegraphics[scale=0.5]{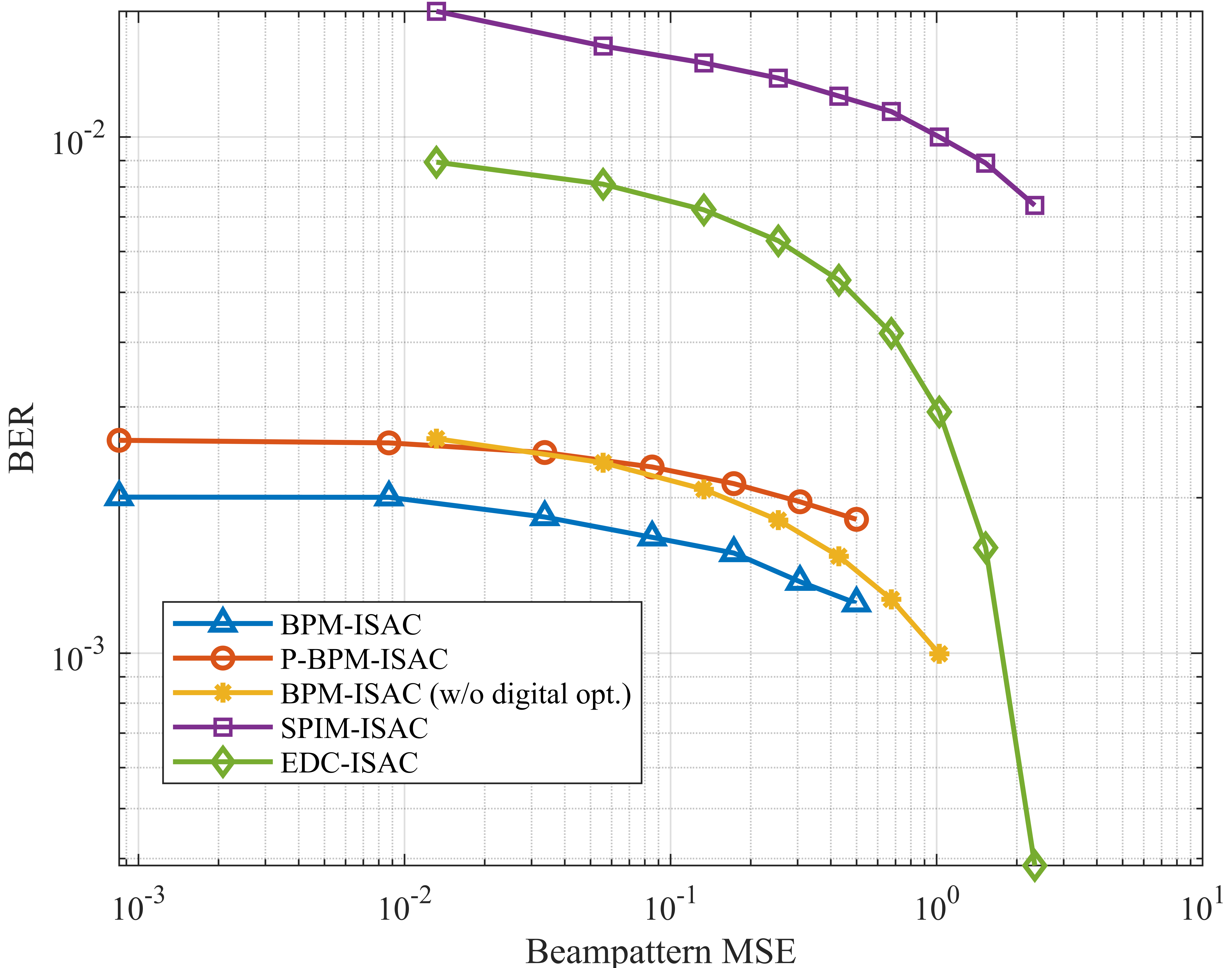}
    \vspace{-2mm}
    \caption{Trade-offs over sensing and communication at $\rm SNR=0$ dB.}
    \label{Trade-off}
    \vspace{-5mm}
\end{figure}
In this section, we consider a fully-connected hybrid mmWave
 ISAC system with $N_{\rm t}=N_{\rm r}=32$.
For communication, we set $P=8$, $N_{\rm C}=3$, $K=4$, and $L=20$.
For sensing, we set average power as $T_{\rm R}=5$ and $W=3$, specifically $\bm{f}(11)$, $\bm{f}(12)$ and $\bm{f}(13)$.
The ideal beampattern is set as  $\bm{t}=\sqrt{T_{\rm R}}\bm{1}_W$ and the activation probability matrix is set as $\bm{D}=\frac{1}{W}\bm{I}_{W}$. 
Additionally, we set the convergence tolerance as $0.001$ for the alternating algorithm. 
The signal-to-noise ratio (SNR) is defined as $\frac{E_b}{N_0}=\frac{N_{\rm C}}{\eta\sigma^2}$.

In Fig. \ref{BER}, we compare the BER performance of BPM-ISAC with other schemes.
SPIM-ISAC \cite{elbir2023millimeter} exhibits high BER at high SNR due to severe sensing interference.
The plain version of BPM-ISAC, denoted as P-BPM-ISAC, utilizes $K$ beams simultaneously without index modulation.
It adopts 4-QAM modulation and is optimized with $\mu=0.5$.
In high SNR region, BPM-ISAC demonstrates lower BER than its plain version, highlighting the superiority of beam pattern modulation.
As $\mu$ increases, strengthening the communication constraint, BPM-ISAC approaches similar BER as that of GBM. Additionally, the BER of BPM-ISAC with the on-grid beams is provided, consistent with APEP analysis.

In Fig. \ref{beampattern}, the instantaneous beampatterns of BPM-ISAC under the same channel realization with different $\mu$ are presented.
Without loss of generality, random $N_{\rm C}$ out of $K$ communication beams and sensing beam $\bm{f}(11)$ are activated for explanation.
As $\mu$ grows, the tolerance towards strong sensing interference raises, and the sensing beampattern approaches the desired one with unoptimized power allocation.

In Fig. \ref{Trade-off}, the trade-off curves between BER and beampattern MSE among different methods are presented.
For EDC-ISAC, eigenvectors corresponding to $K$ largest eigenvalue of the spatial channel are utilized to construct EDC.
The proposed scheme outperforms all others thanks to optimized hybrid transceivers.
In addition, the performance of BPM-ISAC without digital part optimization is given to demonstrate the effectiveness of power allocation.
\vspace{-1mm}
\section{Conclusions}
In this paper, we have proposed a novel IM-embedded ISAC scheme, termed BPM-ISAC, specifically designed for mmWave hybrid structures. 
BPM-ISAC aims to retain the SE benefits of primitive beamspace modulation schemes while addressing performance bottlenecks in their extension to  ISAC functionalities. 
To ensure near-optimal performance for BPM-ISAC, we formulated an optimization problem to minimize the sensing beampattern MSE under the communication MSE constraint and solved it by optimizing analog and digital parts sequentially. 
Theoretical analysis and simulation results verified that the proposed BPM-ISAC outperforms existing candidates in terms of sensing and communication trade-offs.
\vspace{-3mm}
\bibliographystyle{IEEEtran}
\small
\bibliography{IEEEabrv, ref}
\end{document}